\documentclass[10pt]{article}
\usepackage[dvips]{graphicx}

\usepackage{epsfig}
\usepackage{graphicx}
\usepackage{indentfirst}
\usepackage{amsmath}
\usepackage{amsfonts}
\usepackage{amssymb}

\begin{document}

\begin{center}

{\Large\bf SPIRAL GALAXIES ROTATION CURVES\\[5PT]
WITH A LOGARITHMIC CORRECTED\\[5PT]
NEWTONIAN GRAVITATIONAL POTENTIAL\\[5PT]}
\medskip

J.C.
Fabris\footnote{e-mail: fabris@pq.cnpq.br. Present address:
Institut d'Astrophysique de Paris (IAP), 98bis Boulevard Arago,
75014 Paris, France.} and J. Pereira Campos\footnote{e-mail: jpcampospt@gmail.com},
\medskip

Departamento de F\'{\i}sica, Universidade Federal do Esp\'{\i}rito
Santo\\ 29060-900, Vit\'oria, Esp\'{\i}rito Santo, Brazil
\medskip

\end{center}

\begin{abstract}

We analyze the rotation curves of 10 spiral galaxies with a
newtonian potential corrected with an extra logarithmic term,
using a disc modelization for the spiral galaxies. There is a new
constant associated with the extra term in the potential. The
rotation curve of the chosen sample of spiral galaxies is well
reproduced for a given range of the new constant. It is argued
that this correction can have its origin from string
configurations. The compatibility of this correction with local
physics is discussed.

\end{abstract}

Pacs numbers: 98.80.-k, 98.80.Es

The observed features of the rotation curves of gravitationally
bound objects, like stars or gas clouds, in spiral galaxies is one
of the most important evidence to the existence of dark matter.
Spiral galaxies is composed essentially of a very dense central
bulge and a flat disc containing spiral arms. At large distance
from the center of the galaxy, the keplerian orbits of stars or
gas clouds should exhibit velocities decreasing with the distance
to the center. However, observations indicate that the velocities
become constant (or even increase) at very large distances to the
center of the galaxy. This feature seems to be universal, and it
can not be explained by the visible matter present in the galaxy,
neither by an additional amount of invisible baryonic matter
(baryonic dark matter). It is largely accepted today in the
astrophysics community that the only consistent explanation is the
existence of a large amount of non-baryonic dark matter, which
composes a huge halo around the galaxy extending for a distance
many times larger than the visible radius of the galaxy. The
reader is addressed to the reviews articles on the subject for the
evidences for the existence of non-baryonic dark matter and for
the candidates to represent it \cite{bergstrom,silk}. For specific
models concerning the dark matter hypothesis, see reference
\cite{salucci}.
\par
Concerning the candidates to form this "dark halo" around the
spiral (and perhaps even the elliptic) galaxies, those most
frequently quoted are the axion and the neutralinos. Axions are
relics of "grand unified theory" (GUT) phase in the primordial
universe. Axions are particles with small mass, $m_A \sim 5\,eV$,
but with a very small cross section, so that a gas of axions may
be approximated by a pressureless, non-relativistic, fluid in
spite of its small mass. Neutralinos are the less massive,
electrically neutral, stable supersymmetric particles. Both axions
and neutralinos have all the characteristic one can ask for a
non-baryonic dark matter: they do not interact with light, having
at the same time a very small cross section, behaving essentially
as a cold (pressureless) fluid. Cosmologically, they also could
play an important role in the structure formation since they would
decouple very early from the radiative gas that dominates the
matter content of the universe before the matter dominated phase.
\par
In spite of its many success in describing the dynamics of
galaxies, and even cluster of galaxies, the dark matter model has,
for the moment, a major drawback: no one of its candidates comes
from a experimentally tested theory. Axions and neutralinos, in
special, remain theoretical proposals. This fact legitimates the
search of alternatives to the dark matter model. One of them is to
suppose that the ordinary laws of gravity are not valid from
galactic scales on. This alternative is quite attractive since the
proportion of dark matter, with respect to baryonic matter,
required to explain the dynamics of virialized structure like
galaxies and clusters of galaxies increases as the scales of the
structure increases: clusters of galaxies ask for a much higher
proportion than galaxies, for example \cite{cluster}. A recent analysis of
supercluster dynamics suggests the possibility of a new long range force \cite{farrar}. Another
possibility is to keep the gravity theory intact, but to modify
the law of inertia, as it has been done by the MOND theory, which
modifies the Newton second law \cite{milgrom}. This proposal has
been particularly successful in reproducing the rotation curves of
spiral galaxies \cite{sanders}. The application of MOND to
clusters of galaxy is more problematic.
\par
In this work we will opt to investigate the consequences of
changing the newtonian gravity law. This will be done by adding an
extra logarithmic term to the newtonian potential. This has the
evident advantage of giving a constant velocity to the keplerian
trajectories for large values of the radial distance $r$. But, it
has also a theoretical motivation: string-like objects would
contribute with a logarithmic type potential if the energy and the
radial pressure are much larger than the transverse pressure.
Hence a gas of cosmic string would have a different gravitational
behavior compared with a gas of point-like objects. The
relativistic analysis of a gas of string has been carried out in
reference \cite{letelier}. The connection with a logarithmic extra
term in the newtonian potential has been presented in reference
\cite{soleng}. The logarithmic extra term becomes exact when the
strings composing the gas are all radially aligned.
\par We will
compute here a more detail galactic model with the extra
logarithmic potential term, performing also a comparison of
theoretical predictions and observational data for a given sample
of spiral galaxies. Using a flat thin disc approximation for
representing the spiral galaxies, we will show that the
logarithmic modification leads to very good results. We remark,
{\it en passant}, that a relativistic non-linear theory of the
type $f(R)$, $R$ being the Ricci scalar, leads to a modification
of the newtonian potential with an extra term of the type
$r^\beta$. The comparison for such kind of models (not including a
logarithmic potential term) with the rotation curve of $LSB$
spiral galaxies has been performed in references
\cite{troisi,salucci}. A preferred value $\beta \sim 1.7$ has been
found \cite{salucci}. Here we will use a sample of 9 $LSB$
galaxies described in reference \cite{blok}. To this sample, we
add a galaxy from reference \cite{salucci} whose rotation curve
extend very far from the luminous disk.
\par
Let us consider the modified newtonian potential as
\begin{equation}
\phi(\vec r) = - \frac{Gm}{r} - \alpha
Gm\ln\biggr(\frac{r}{r_0}\biggl) \quad ,
\end{equation}
where $\alpha$ is a new parameter with dimension of $L^{-1}$ (which must be negative in order to give an attractive extra term in the
newtonian force) and $r_0$ is an arbitrary constant. We will treat the new (dimensional) constant $\alpha$ as arbitrary, even if it
must have a connection with the string tension. Considering a smooth distribution of mass,
the potential at a position $\vec r$ is given by
\begin{equation}
\phi(\vec r) = - G\int_{V'}\frac{\rho(\vec r')}{|\vec r - \vec
r'|}d\vec r' - \alpha G\int_{V'}\rho(\vec r')\ln\biggr(\frac{|\vec
r - \vec r'|}{r_0}\biggl)d\vec r' \quad .
\end{equation}
Let us consider now the approximation that a spiral galaxy is a flat disc of a given radius. In this first approach, the bulge and the
spiral arms will be neglected. Hence, we must solve the integrals above in cylindrical coordinates. The disc is characterized by a surface density.
We will use the exponential model for the disc \cite{freeman}, and we will suppose that light traces mass. Hence, the surface density is
given by
\begin{equation}
\Sigma(r) = \Sigma_0e^{- \frac{r}{r_d}} \quad ,
\end{equation}
where $r_d$ is the luminosity scale of the disc, indicating how surface luminosity decreases with the distance to the center of the disc,
and $r$ is the distance of the center of the disc to a given point on it.
\par
For the pure newtonian potential, this calculation is present in some detail in reference \cite{binney}. Adding the calculation to the extra
term, which is quite straightforward, we find
\begin{equation}
\label{cv} v_c^2 = \frac{\pi G\Sigma_0
r^2}{r_d}\biggr[I_0\biggr(\frac{r}{2r_d}\biggl)K_0\biggr(\frac{r}{2r_d}\biggl)
-
I_1\biggr(\frac{r}{2r_d}\biggl)K_1\biggr(\frac{r}{2r_d}\biggl)\biggl]
+ 2\pi\alpha G\Sigma_0 r_d\biggr[e^{-\frac{r}{r_d}}(r + r_d) -
r_d\biggl] \quad ,
\end{equation}
where $I_\nu$ and $K_\nu$ are the modified Bessel's functions.
\par
The first term in (\ref{cv}) represents the contribution of the
newtonian potential, while the second one is the contribution due
to the extra logarithmic term. In relation (\ref{cv}) there are
essentially two free parameters: the central surface density
$\Sigma_0$ and $\alpha$. We will use a sample of nine low surface
brightness galaxies ($LSB$), that is, galaxies that ask for a
large amount of dark matter, to test this model.
\par
The comparison between the observational data and the theoretical
results is given through the $\chi^2$ statistics:
\begin{equation}
\chi^2 = \sum_i\frac{(v_i^o - v_i^t)^2}{\sigma_i^2} \quad ,
\end{equation}
where $v_i^o$ is the observed velocity of the $ith$ object,
$v_i^t$ the theoretical result for this same object, and
$\sigma_i$ the observational error bar. From this, we obtain the
probability distribution
\begin{equation}
P(\alpha,\Sigma_0) = Ae^{-\chi^2(\alpha,\Sigma_0)/2} \quad ,
\end{equation}
$A$ being a normalization constant. The prediction for a given
parameter is obtained by integrating (marginalizing) on the other
parameter, resulting in one-dimensional probability
distribution. More details on this procedure can be found in
reference \cite{colistete1,colistete2}.
\begin{center}
\begin{figure}[!t]
\begin{minipage}[t]{0.225\linewidth}
\includegraphics[width=\linewidth]{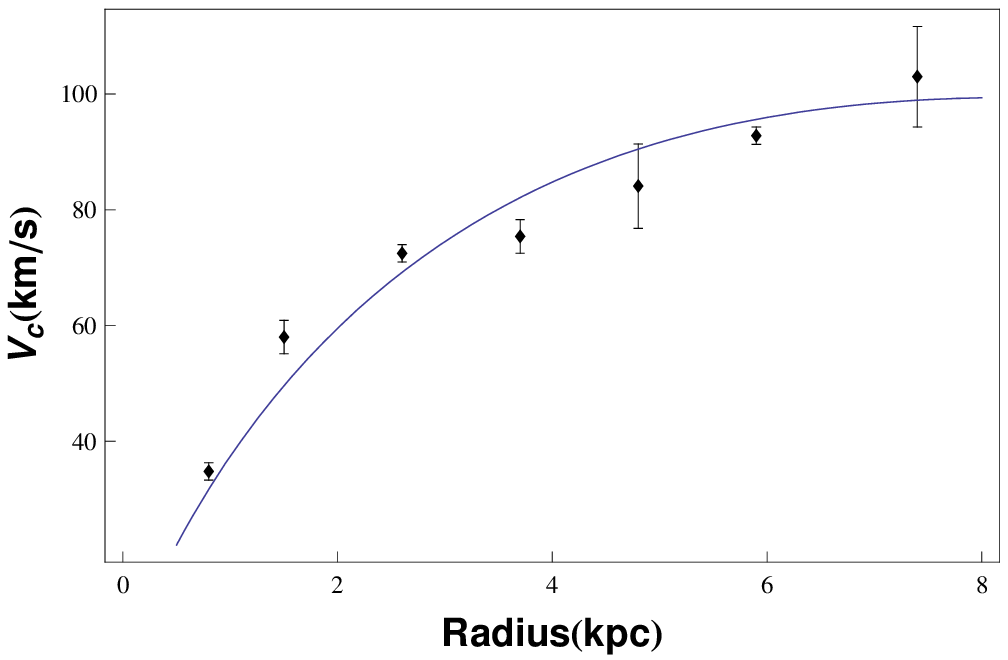}
\end{minipage} \hfill
\begin{minipage}[t]{0.225\linewidth}
\includegraphics[width=\linewidth]{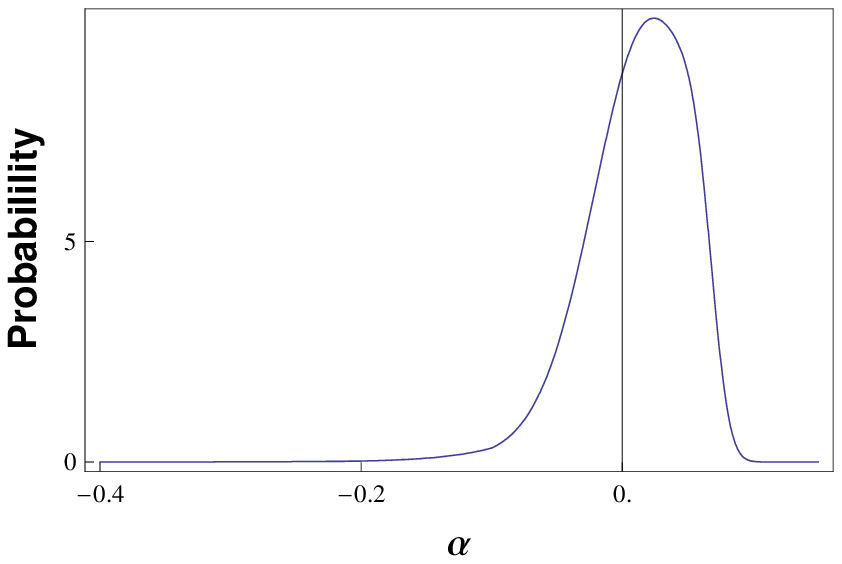}
\end{minipage} \hfill
\begin{minipage}[t]{0.225\linewidth}
\includegraphics[width=\linewidth]{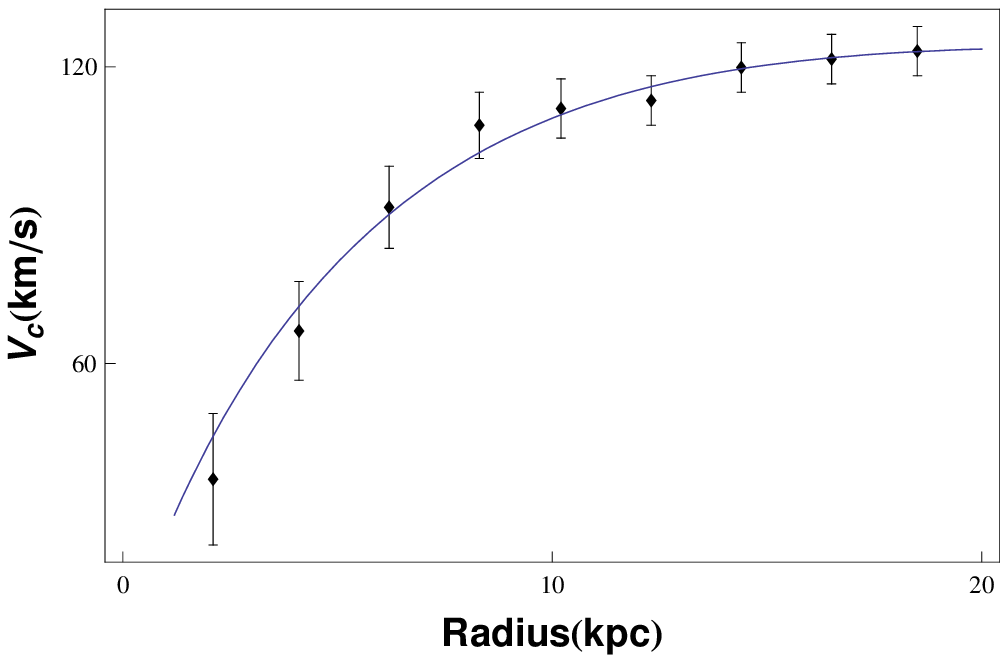}
\end{minipage} \hfill
\begin{minipage}[t]{0.225\linewidth}
\includegraphics[width=\linewidth]{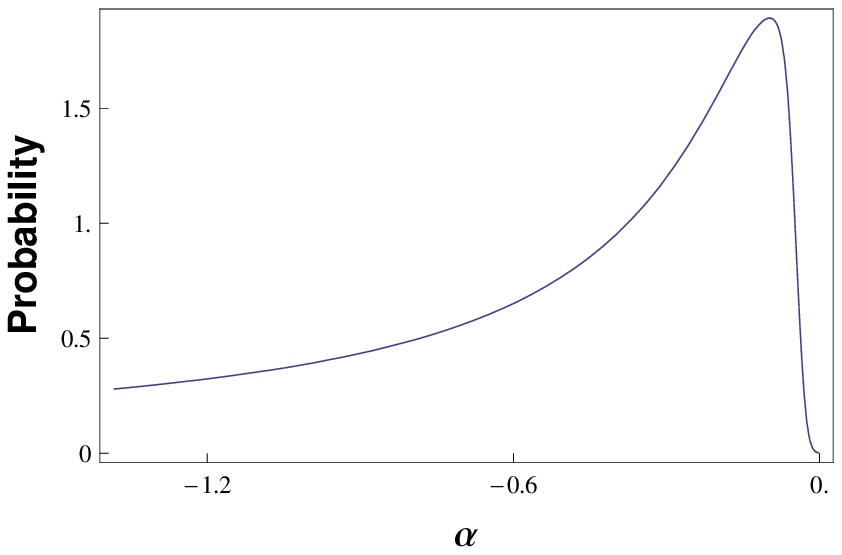}
\end{minipage} \hfill
\begin{minipage}[t]{0.225\linewidth}
\includegraphics[width=\linewidth]{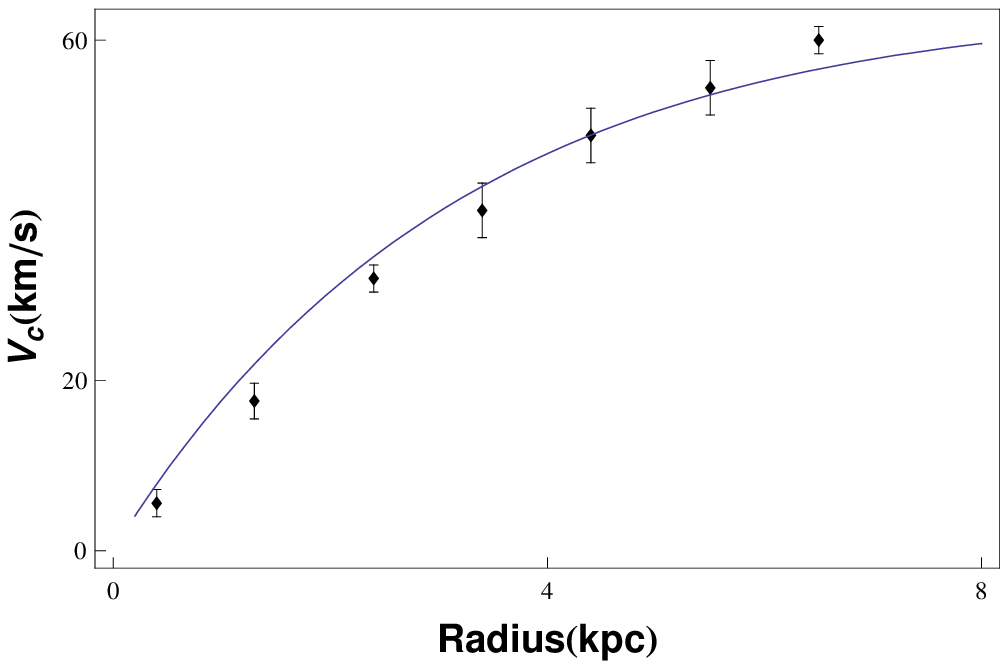}
\end{minipage} \hfill
\begin{minipage}[t]{0.225\linewidth}
\includegraphics[width=\linewidth]{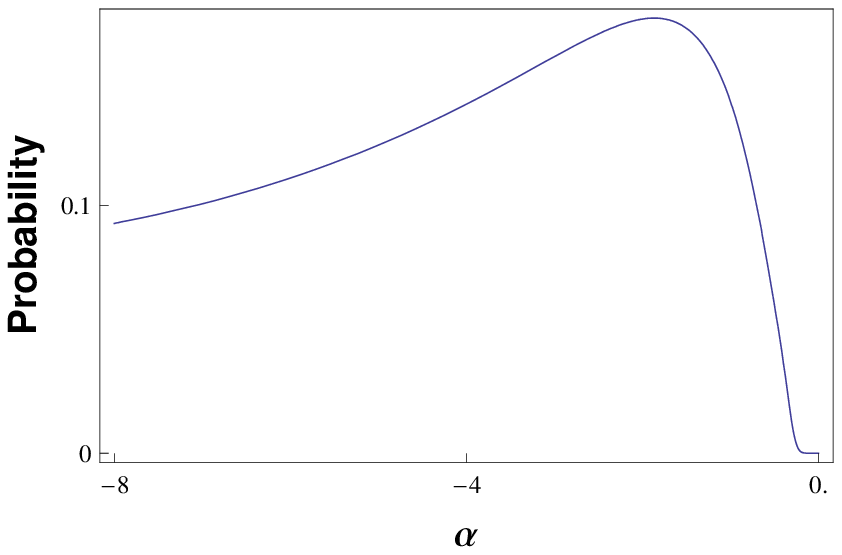}
\end{minipage} \hfill
\begin{minipage}[t]{0.225\linewidth}
\includegraphics[width=\linewidth]{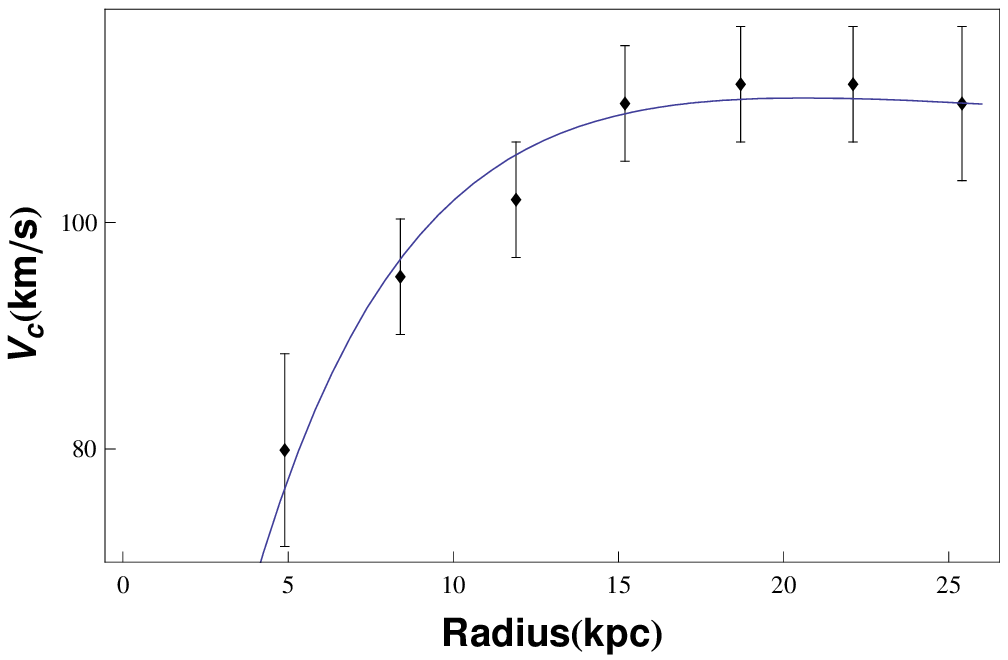}
\end{minipage} \hfill
\begin{minipage}[t]{0.225\linewidth}
\includegraphics[width=\linewidth]{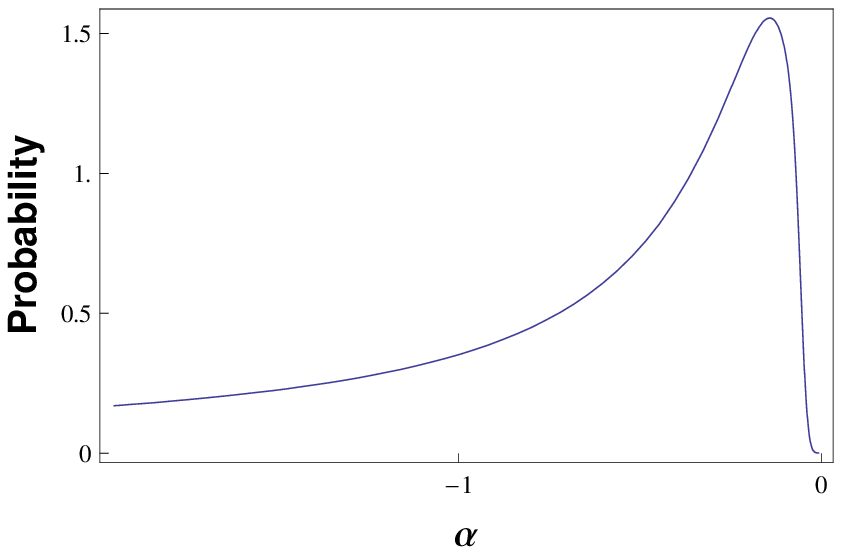}
\end{minipage} \hfill
\begin{minipage}[t]{0.225\linewidth}
\includegraphics[width=\linewidth]{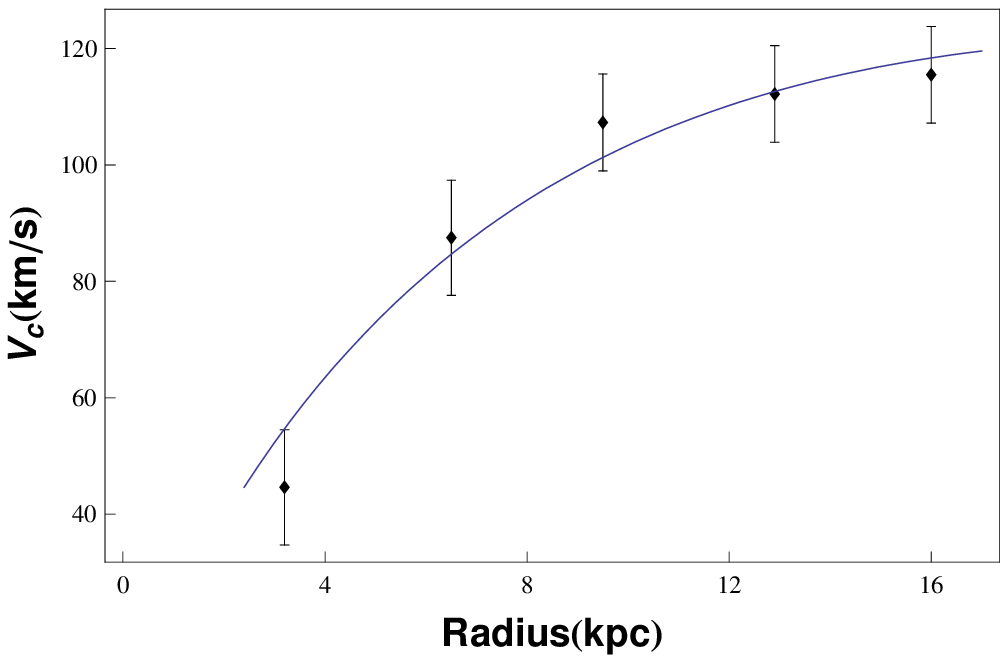}
\end{minipage} \hfill
\begin{minipage}[t]{0.225\linewidth}
\includegraphics[width=\linewidth]{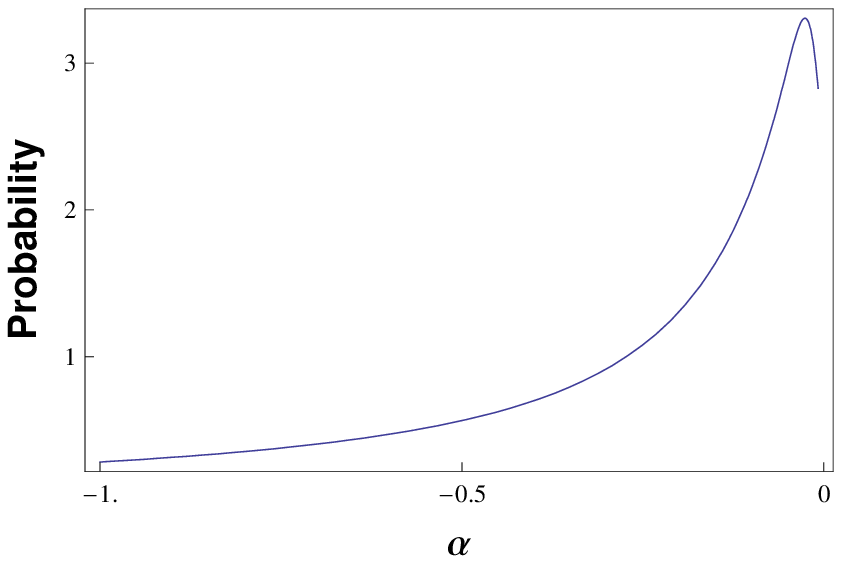}
\end{minipage} \hfill
\begin{minipage}[t]{0.225\linewidth}
\includegraphics[width=\linewidth]{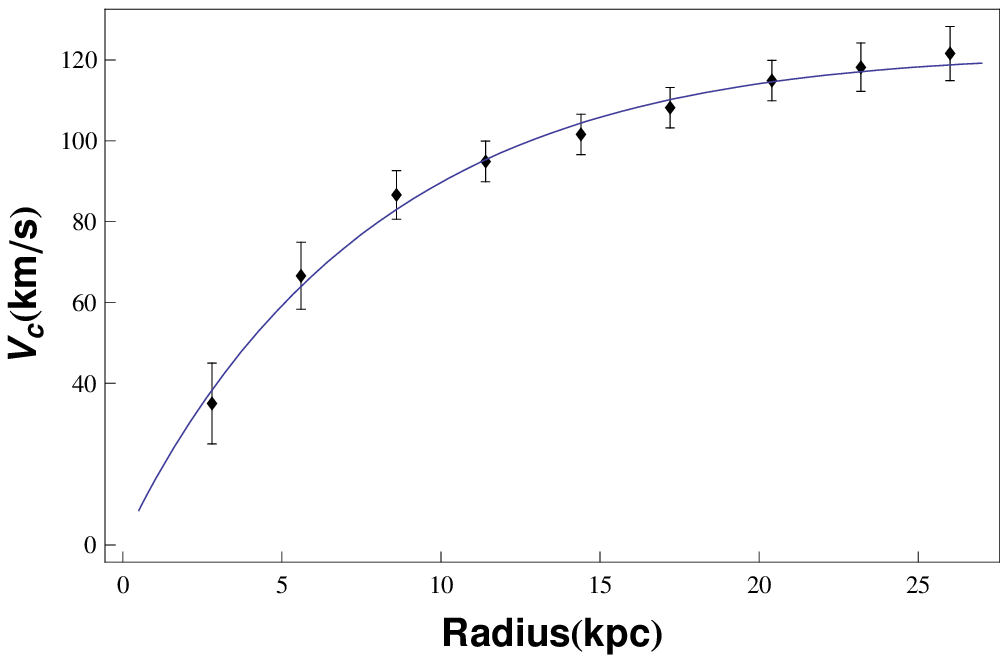}
\end{minipage} \hfill
\begin{minipage}[t]{0.225\linewidth}
\includegraphics[width=\linewidth]{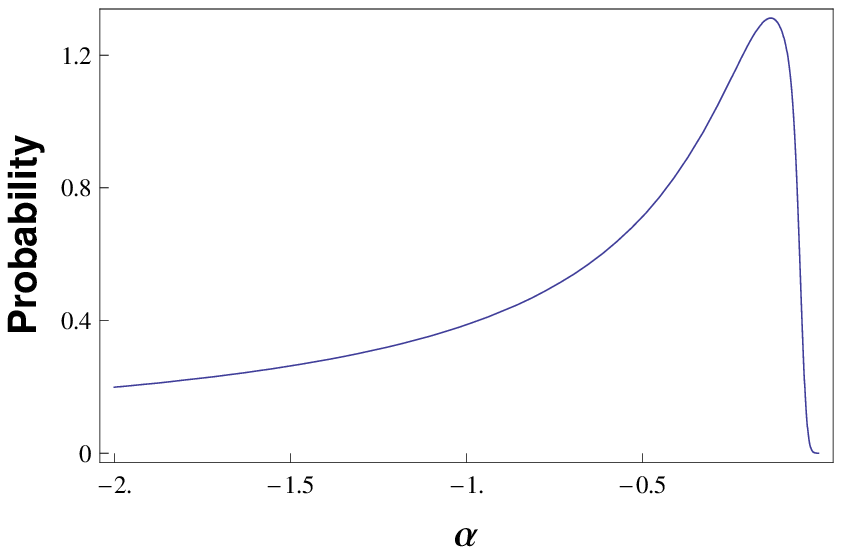}
\end{minipage} \hfill
\begin{minipage}[t]{0.225\linewidth}
\includegraphics[width=\linewidth]{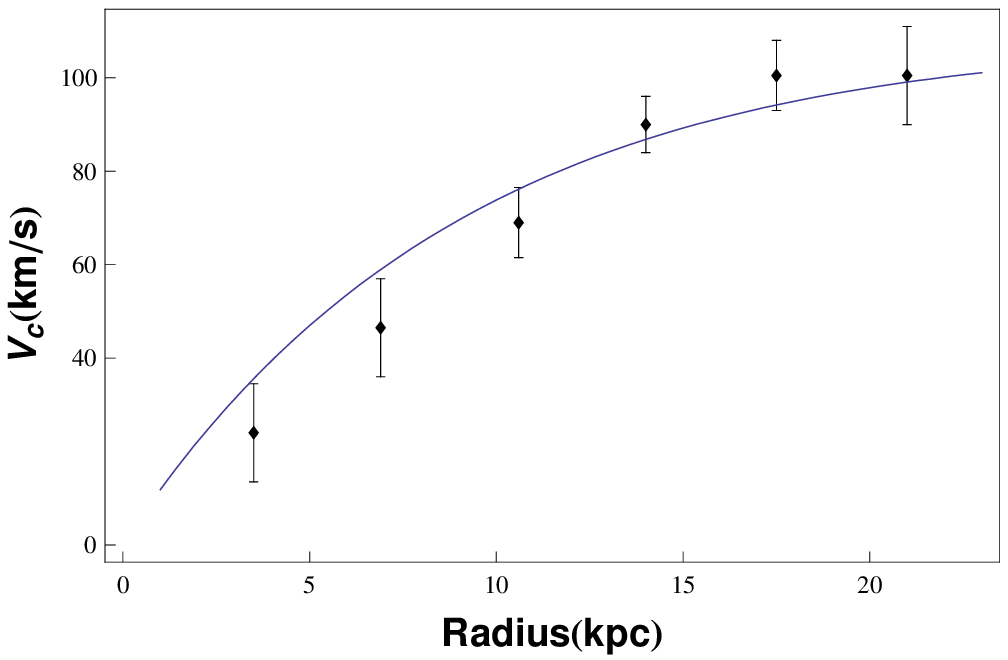}
\end{minipage} \hfill
\begin{minipage}[t]{0.225\linewidth}
\includegraphics[width=\linewidth]{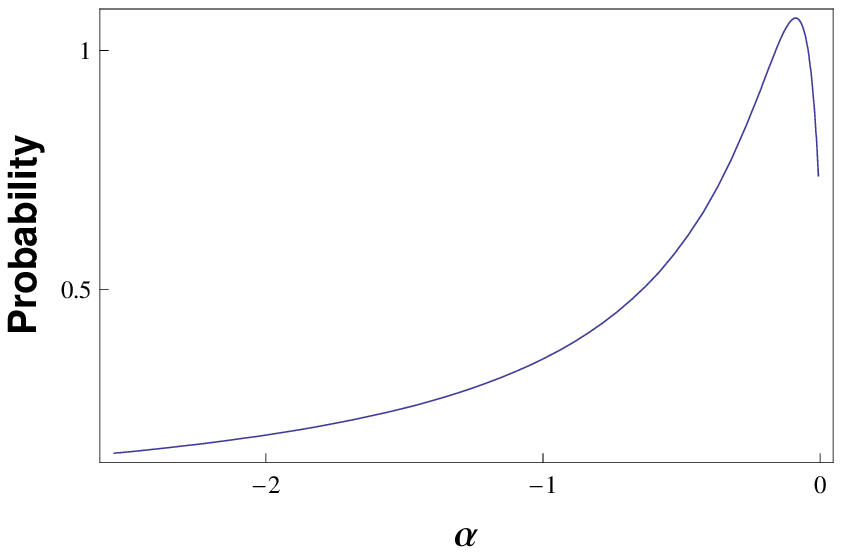}
\end{minipage} \hfill
\begin{minipage}[t]{0.225\linewidth}
\includegraphics[width=\linewidth]{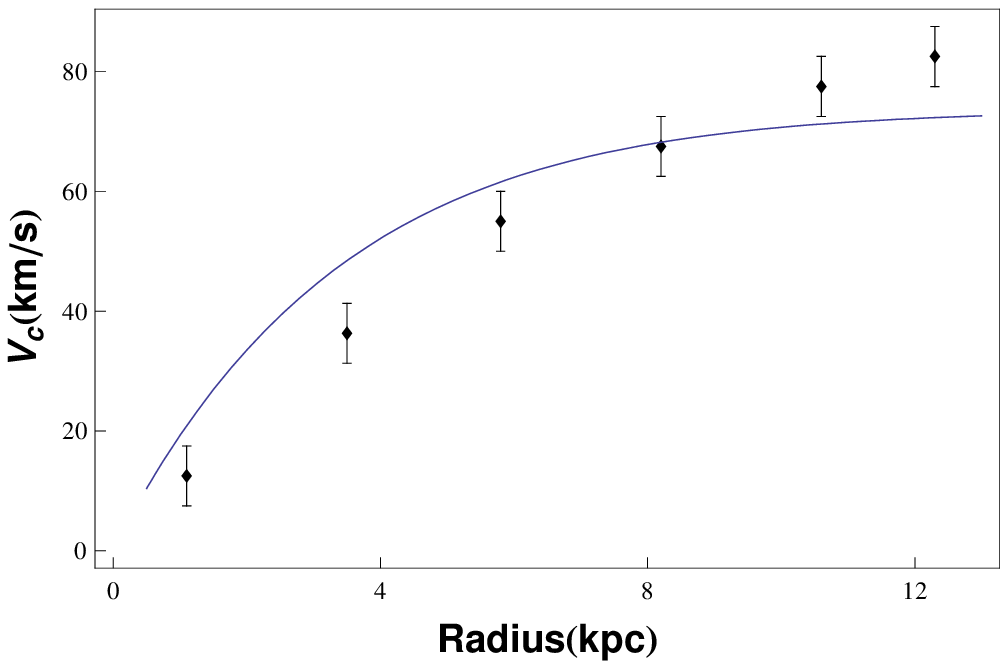}
\end{minipage} \hfill
\begin{minipage}[t]{0.225\linewidth}
\includegraphics[width=\linewidth]{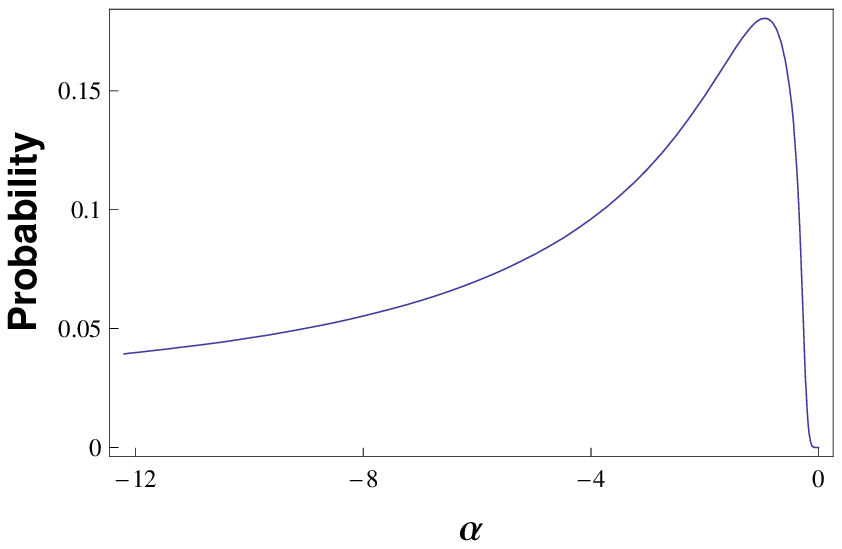}
\end{minipage} \hfill
\begin{minipage}[t]{0.225\linewidth}
\includegraphics[width=\linewidth]{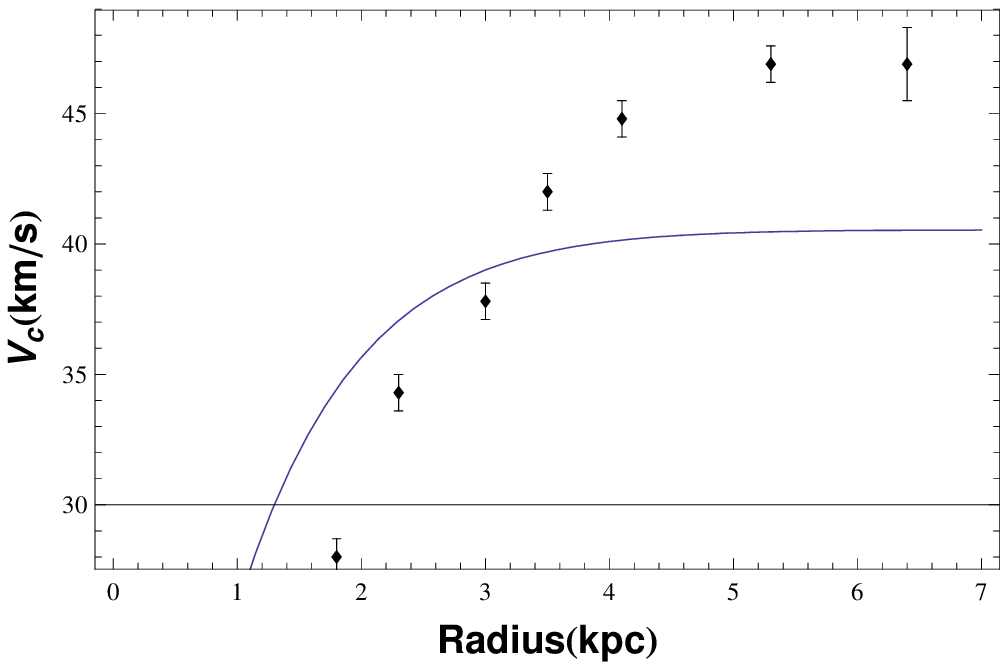}
\end{minipage} \hfill
\begin{minipage}[t]{0.225\linewidth}
\includegraphics[width=\linewidth]{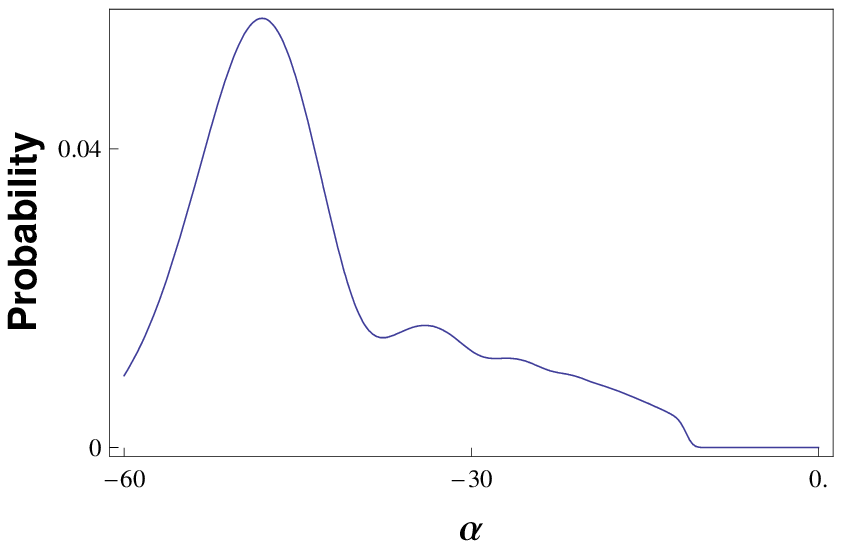}
\end{minipage}\hfill
\begin{minipage}[t]{0.225\linewidth}
\includegraphics[width=\linewidth]{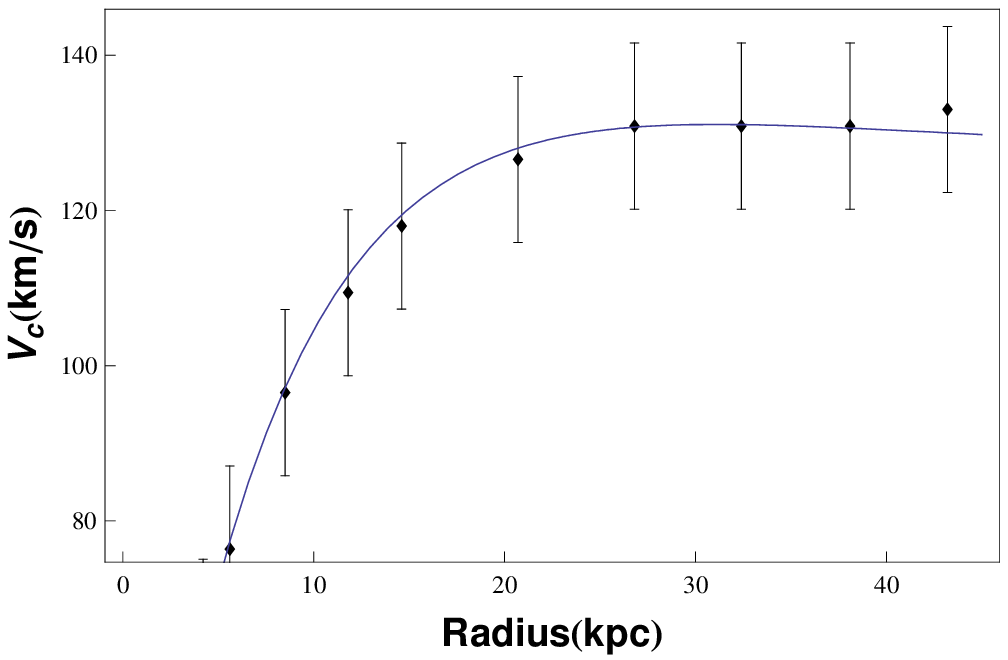}
\end{minipage}\hfill
\begin{minipage}[t]{0.225\linewidth}
\includegraphics[width=\linewidth]{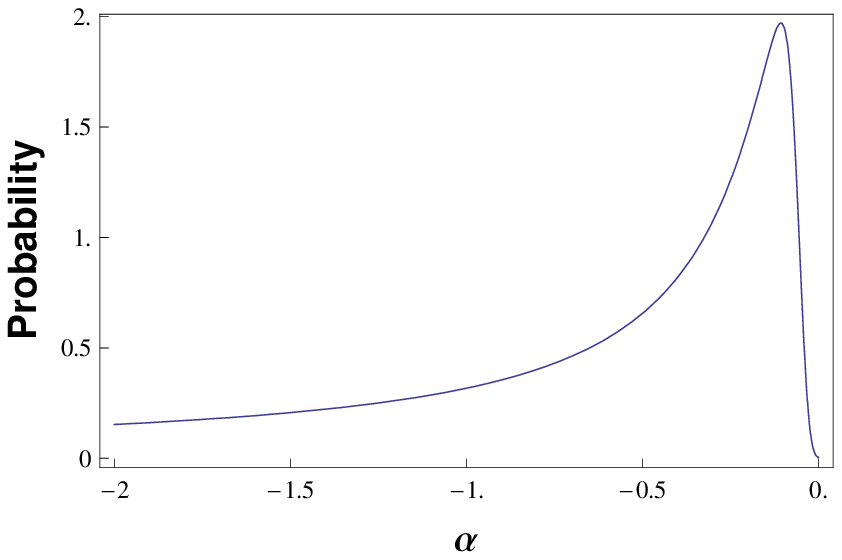}
\end{minipage} \hfill
\caption{{\protect\footnotesize The rotation curve for the $LSB$
spiral galaxies $NGC247$, $F568.v1$, $NGC3109$, $F563-1$,
$F568-1$, $F568-3$, $F574-1$, $F583-1$, $DDO154$ and $UGC128$. The
continuous curve is the theoretical results using the modified
newtonian potential. In the right of each rotation curve it is
displayed the corresponding marginalized probability distribution
for the parameter $\alpha$.}}
\end{figure}
\end{center}

\begin{center}
\begin{tabular}{|c|c|c|}
\hline Galaxy&$\alpha (2\sigma)$&$r_d$\\ \hline
NGC247&$- 0.0243^{+0.0469}_{-0.1062}\,kpc^{-1}$& $2.30\,kpc$\\
\hline F568.v1&$- 0.
0975^{+0.0911}_{-1.1293}\,kpc^{-1}$&$4.86\,kpc$\\ \hline
NGC3109&$- 1.8643^{+1.1910}_{-5.8406}\,kpc^{-1}$&$2.30\,kpc$\\
\hline F563-1&$- 0.1421^{+0.1181}_{-1.5390}\,kpc^{-1}$ &
$4.29\,kpc$\\ \hline F568-1&$
-0.0261^{+0.0179}_{-1.2871}\,kpc^{-1}$&$8.20\,kpc$\\ \hline
F568-3&$- 0.1353^{+0.1304}_{-1.3127}\,kpc^{-1}$&$6.15\,kpc$\\
\hline F574-1&$-
0.0883^{+0.0866}_{-2.1433}\,kpc^{-1}$&$6.58\,kpc$\\ \hline F583-1&
$- 0.9462^{+0.6937}_{-10.0875}\,kpc^{-1}$&$2.43\,kpc$\\ \hline
DDO154&$- 48.0835^{+15.9865}_{-3.8165}\,kpc^{-1}$&$0.72\,kpc$\\
\hline UGC128&$-0.1065^{+0.0795}_{-1.6054}\,kpc^{-1}$&$6.4\,kpc$\\
\hline
\end{tabular}
\end{center}

\par
In the sample of 10 galaxies, two of them are dwarf and one is a
barred irregular. From figure (1) and from the corresponding
table, we can verify that the results are very good, giving a
consistent value for the parameter $\alpha$ for the "normal" $LSB$
spiral galaxies. For the two dwarfs galaxies the results give
either a very small value to $\alpha$ ($NGC247$) or a too large
value ($NGC3109$). Moreover, the results are clearly inconsistent
for the irregular barred galaxy $DDO154$. These three cases
suffer, in fact, from the simplicity of the model: these galaxies,
and mainly the $DDO154$, are poorly approximated by a flat disc.
For the other six, however, the approximation works remarkably
well. For these six galaxies, the predicted value of $\alpha$ is
always negative (at $2\sigma$ confidence level), indicating an
attractive extra term. There is a "concordance" value around
$\alpha \sim - 0.1\,kpc^{-1}$.
\par
A correction of the newtonian potential with $\alpha \sim -
0.1\,kpc^{-1}$ may lead to the "good" rotation curves for spiral
galaxies. At the same time, it does not imply a detectable effect
at level of the solar system. In fact, the ratio between the
normal newtonian force $F_N$ and the correction to this force
$F_c$ is of the order of
\begin{equation}
\frac{F_c}{F_N} = - \alpha r \quad .
\end{equation}
For a distance of some solar units, and using the indicated value
for $\alpha$, this ratio is of the order $10^{-13}$. The extra
acceleration it provokes in the solar system, at a distance of
some solar units from the sun, is of the order $10^{-12}$: the
Pioneer anomaly is surely not due to this correction. This is
reasonable, since a gas of cosmic strings should not have
detectable effects at the solar system scales, otherwise cosmic
stings should already been observed in local experiences.
\par
The most important point to be stressed is the fact that the value
of $\alpha$ is compatible for the group of galaxies for which the
disc approximation is at least reasonable. For the dwarf galaxies
this approximation is quite crude. For the irregular barred galaxy
$DDO154$, the approximation is simply invalid. Hence, even the
fact that the results for this last case is inconsistent with the
other cases seems to be a positive aspect of the model exploited
here, due to the simplicity of the modelization made. Moreover, we
remark the excellent agreement with the galaxy $UGC128$ whose
observed rotation curve extend very far from the luminous disk.
\newline
 \vspace{0.5cm}
 \newline
 {\bf Acknowledgements}:\\
 We thank CNPq (Brazil) and the french/brazilian
scientific cooperation CAPES/COFECUB (project number 506/05) for
partial financial support. We thank also Winfried Zimdahl for reading the text and his suggestions.

\end{document}